\documentclass[useAMS,usenatbib]{mn2e}

\usepackage{graphicx}  

\def\simgt{\lower.5ex\hbox{$\; \buildrel > \over \sim \;$}}
\def\simlt{\lower.5ex\hbox{$\; \buildrel < \over \sim \;$}}

\def\msun{$M_\odot$}
\def\Teff{$T_{\rm eff}$}
\def\teff{$T_{\rm eff}$}

\def\alfamlt{$\alpha_{\rm MLT}$}

\def\alfa2d{MLT--$\,\alpha^{\rm 2D}$}

\def\taufot{$\tau_{\rm ph}$}
\def\logg{$\log g$}

\title[New light on the old problem of lithium pre--MS depletion]{New light on 
the old problem of lithium pre--MS depletion: models with 2D RHD convection}
\author[J. Montalb\'an and F. D'Antona]{J. Montalb\'an$^{1}$\thanks{E-mail: 
j.montalban@ulg.ac.be (JM); dantona@oa-roma.inaf.it (FD)} and F. D'Antona$^{2}$\\
$^{1}$Universit\'e de Li\`ege, Institut d'Astrophysique et G\'eophysique, 
    All\'ee du 6 A\^{out},  B-4000 Li\`ege, Belgium\\
$^{2}$INAF, Osservatorio di Roma, I-00040 Monteporzio, Italy}
\begin{document}

\date{}
\pagerange{\pageref{firstpage}--\pageref{lastpage}} \pubyear{2006}

\maketitle

\label{firstpage}

\begin{abstract}

The \teff\ location of Pre-Main Sequence (PMS) evolutionary tracks depends on 
the treatment of over-adiabaticity. We present here the  PMS evolutionary tracks 
computed by using the mixing length theory of convection (MLT) in which the 
$\alpha_{\rm MLT}=l/H_p$ parameter calibration is based on 2D--hydrodynamical models 
(Ludwig et al. 1999). These MLT--$\,\alpha^{\rm 2D}$ stellar models and tracks are very 
similar to those computed with  non--grey 
ATLAS9 atmospheric boundary conditions and  Full Spectrum of Turbulence 
(FST) convection model both in the atmosphere and in the interior. 
The comparison of the new tracks with  the location on the HR diagram  of pre--MS binaries
is not completely satisfactory, as some binary components are located at too low \teff. 
Besides, the pre--MS lithium depletion in the MLT--$\,\alpha^{\rm 2D}$  tracks is  still much 
larger than  that expected from the observations of lithium   
in  young open clusters. This result is similar to that of FST models. 
Thus, in spite of  the fact that 2D RHD 
models should provide a better  convection description  than any local 
model, their introduction is not sufficient to reconcile theory and observations.
Lithium depletion in young clusters points towards a convection 
efficiency which, in pre--MS, should be  smaller than in the MS. 
The pre--MS 
lithium depletion  decreases significantly  in FST models if we reduce the 
solar metallicity down to the value suggested by Asplund et al.~(2004), but the 
corresponding solar model does not reproduce  the depth of the convective zone as 
determined by helioseismology.

\end{abstract}


\section{Introduction}

The location in the Hertzsprung--Russell diagram  (HRD)  of Pre-Main Sequence (PMS) 
 evolutionary tracks is very sensitive to modeling details  
 such as  opacity, atmospheric boundary conditions, rotation and 
convection treatment. \cite{josi2004} and \cite{dm2003} pointed out that,  at 
least for \teff\ down to 4000~K, the treatment of convection transport plays a 
role on the PMS evolutionary tracks that is  much more important  than the 
effects from recent improvements of low temperature opacities. They showed that 
it is the ``average efficiency" of convection in the envelope which determines 
both, the \teff\ of PMS tracks and the lithium depletion\footnote{This 
conclusion is valid only if  no other transport process different than 
convection occurs during the PMS.} during this phase of stellar evolution.

Unfortunately, the approaches traditionally used in stellar evolution 
to describe the heat transport by convection are quite rough.
The  mixing length theory (MLT, B\"ohm-Vitense 1958) as well as the 
Full Spectrum of Turbulence (FST) model \citep{cm,cgm}
 are local theories, which ignore physical processes such as convective overshoot 
and radiative transfer effects.

The efficiency of convection in the superadiabatic region at the top of convective 
stellar envelopes determines the asymptotic value of the entropy in the deep and  
adiabatically stratified layers and, therefore, the star radius  
and \teff. 
Any treatment of convection, adjusted to obtain the correct adiabat, 
will provide similar global structures. This is the base of the MLT solar 
calibration.  Given the relationship between convective flux and over--adiabaticity in 
the MLT ($F_{\rm conv}\propto (\nabla-\nabla _{\rm ad})^{3/2}\alpha_{\rm MLT}^{2}$), 
increasing the value of the $\alpha_{\rm MLT}$ parameter  we decrease the 
value of the over-adiabaticity, and the radius decreases.

The only parameter in the MLT, $\alpha_{\rm MLT}$,  is tuned to reproduce the solar 
radius at the solar age. This method has provided very good results, in spite of  its simplicity. 
 Nevertheless, the Sun yields only one calibration value, 
and no physical principle guarantees  this value to be appropriate for the whole 
HR diagram, or  the whole stellar structure. 

The mass of the stellar convective envelope increases with decreasing 
\teff, and its top gradually shifts towards larger depths, so that the 
structure of the surface layers is less and less affected by  convection as 
\teff\ decreases. For temperatures lower than 4700~K, however, the convection 
zone rises again due to H$_2$ dissociation. Furthermore, as gravity 
decreases, the convective flux decreases because of the lower density, and 
therefore, the over-adiabatic region is more and more extended in the stellar 
atmosphere. Specific atmosphere models with an adequate treatment of radiation and 
convection transports are hence mandatory. 

The available grids of model atmospheres for general use of stellar structure  
(ATLAS9 by Kurucz 1993 and NextGen by Hauschildt et al.~1999) were 
generally computed just for one specific convection model, e.g. a given 
$\alpha_{\rm MLT}=\alpha_{\rm MLT}^{\rm atm}$ value in the MLT. Hence, in the solar calibration, 
only the MLT parameter used in the interior can be changed. \cite{josi2004} 
showed that this kind of procedure can lead to an uncertainty of the order of 
200~K in the 1~\msun\ PMS track (without changing the effective temperature of 
MS), since its \teff\ depends on the optical depth chosen to match the interior 
and the atmosphere models.
On the other hand, Heiter et al.~(2002)  computed  FST  ATLAS9 models
with a value of the $\alpha_{\rm FST}$ parameter fixed by  a grey FST solar calibration.
A non--grey FST solar calibration requires a  larger value of $\alpha_{\rm FST}$
(0.18 in the atmosphere and in the interior --Ventura, private communication). Nevertheless, 
given the different properties of $\alpha_{\rm FST}$ and $\alpha_{\rm MLT}$ parameters, 
 a discrepancy between   $\alpha_{\rm FST}^{\rm atm}$
and $\alpha_{\rm FST}^{\rm in}$ have no very serious consequences on the stellar structure
or on the evolutionary tracks (as shown in \cite{josi2004}, this discrepancy 
can lead to an  uncertainty in \teff --at given mass and
luminosity-- of the order of 2\%, and $<$1\% for 1~\msun).
In any case,  non--grey FST stellar models  (either with  $\alpha_{\rm FST}^{\rm in}$=$\alpha_{\rm FST}^{\rm atm}$,
or with $\alpha_{\rm FST}^{\rm in}$=2$\cdot\alpha_{\rm FST}^{\rm atm}$) predict
a too large pre-MS lithium depletion compared to observational data in open clusters.

A more realistic approach consists in solving the hydrodynamic equations coupled to the 
equation of radiative transfer. Recently much progress has been made in 
performing 2-3D radiative--hydrodynamical (RHD) simulations of stellar surface 
convection (e.g. Freytag et al. 1996; Stein \& Nordlund 1998; Asplund et al. 
2000; Ludwig et al. 2002). These results, however, cannot be directly used in 
stellar evolution computations. A way of overcoming the problem of computing 
convection would be to have grids of 3D non local models, and, at each (T$_{\rm 
eff}$-gravity), calibrate the $\alpha$\ value  which provides the same specific 
entropy jump between the atmosphere and the adiabatic region.  This procedure 
does not give information on the structure of the overadiabatic layers, but 
allows a proper computation of the interior\footnote{The use of an ``average 
 efficiency" of envelope convection is then useful if we wish to obtain general 
structural information concerning the stellar interior properties, exactly such 
as the lithium depletion, and the evolution of \teff\ and gravity. Of course, 
it is less useful if we need to know the structure of the overadiabatic layers, 
e.g. for the computation of acoustic oscillation modes.}. 
 
Extensive 3D RHD simulations for general applications to the computation of 
stellar evolution are not yet available, but 2D RHD simulations have been 
performed by \cite{ludwig1999} for a large grid of \teff\ and gravity values, 
and the structural information  from these models has been translated 
into an effective mixing-length parameter $\alpha_{\rm MLT}^{\rm 2D}$, 
suitable to construct standard stellar structure models. 
 
 As the results from RHD models are in principle more reliable than the local 
 models, we computed pre--MS and MS evolution, based on \cite{ludwig1999}, for 
 masses from 0.8 to 1.5~\msun.  We compare the resulting evolutionary tracks and 
 Li abundance with those from  recent local 
 models \citep{josi2004}, which assume FST convection both in the non--grey 
 atmosphere and in the envelope structure.

The application of 3D RHD atmosphere computation to studies of  spectral line 
formation, have also shown that in many cases standard 1D analyses are very 
misleading in terms of derived element abundances (Asplund 2005). 
In particular,  3D  computations together with a much more accurate analysis
 of solar spectrum and 
a non-LTE treatment,  led to a drastic revision of the photospheric solar 
composition. Asplund et al.~(2004 and 2005) analysis results in a significant 
reduction of   solar metallicity down to $Z/X\sim 0.017-0.018$. We expect that 
a smaller lithium depletion could be obtained if we revise downward the solar 
metallicity. Unfortunately, the 2D RHD models by Ludwig et al. (1999), and the 
corresponding $\alpha_{\rm MLT}^{\rm 2D}$ calibration, are only available  
for the higher  ``old" solar 
metallicity and helium mass fraction Y=0.28. As we show in section~3 that 
FST and MLT--$\,\alpha^{\rm 2D}$ results are very similar for the standard ``old"
solar metallicity,  we compute FST pre--MS tracks with 
the scaled down metal abundances, to test if the new abundances could solve 
the problem of PMS-lithium depletion in young open clusters.

\section{``Average efficiency'' of convection for pre--MS models}

\cite{ludwig1999}, by using their 2D RHD atmosphere models,
 provided a calibration of the $\alpha_{\rm MLT}$ parameter 
as a function of $T_{\rm eff}$ and $\log g$ in the domain $T_{\rm eff} = 4300 - 7100$~K, 
$\log g = 2.54 - 4.74$. We would like to stress two interesting aspects of this calibration in the 
PMS evolution context: first, the 2D models 
indicate that convection is on average `efficient' in the atmosphere and the
envelope, corresponding to a large $\alpha_{\rm MLT}$ value; and, second,  
$\alpha_{\rm MLT}$ increases as $T_{\rm eff}$  decreases.
The $\alpha_{\rm MLT}^{\rm 2D}$ behavior has been confirmed by 3D atmosphere models.
 Ludwig, Allard and Hauschildt (2002) find, for an M dwarf at 
$T_{\rm eff} = 2800$~K and $\log g =5$,   \alfamlt$\simeq$2.1, a value noticeably larger
than  that required for the Sun (\alfamlt$\simeq$1.6).  
\cite{trampedach} made a similar $\alpha_{\rm MLT}^{\rm 3D}$ calibration by using 
 3D model atmospheres computed by means of a different code \citep[the time--dependent, 
compressible, explicit, radiative-hydrodynamics code by][]{stein1998}.  
For the range of main sequence gravities  and  $\log T_{\rm 
eff}$=3.68--3.83, this $\alpha_{\rm MLT}^{\rm 3D}$ calibration shows the same behavior
than the $\alpha_{\rm MLT}^{\rm 2D}$ one and
a systematic offset, in the sense $\alpha_{\rm MLT}^{\rm 3D} > \alpha_{\rm MLT}^{\rm 2D}$. 
For the Sun, \cite{ludwig1999} found  than  their  2D-based calibration
underestimates the value of $\alpha_{\rm MLT}$ by a factor $\sim 0.2$, and  they explained 
that as due to a combined effect of low-temperature opacities and the 2D approximation.
On the  other hand, \cite{asplundsol2000} have compared 2D and 3D atmosphere models for 
Sun, and find that  the 2D solar model  has marginally larger gradients than 
the 3D one. 
\cite{ludwig1999} propose the use of a constant scaling factor to compensate  the systematic
offset seen in the Sun, as the relative variations of $\alpha_{\rm MLT}$ with \teff\ and \logg\ should
be less affected by systematic shortcomings such as the 2D approximation.

It has also been  shown that there is no unique value of the $\alpha_{\rm MLT}$ parameter 
that can  reproduce the temperature gradients in the over-adiabatic regions 
\citep{steffen1999,trampedach2004}. Nevertheless, the function  
\alfamlt(\Teff,\logg)  by \cite{ludwig1999}, guarantees that the 
adiabat of a stellar model, computed with this $\alpha_{\rm MLT}^{\rm 2D}$  and grey boundary 
conditions,  provides the same adiabat (and therefore the same radius and 
effective temperature) as one from a complete 2D RHD computation.

\begin{figure}
\begin{center}
\includegraphics[width=.48\textwidth]{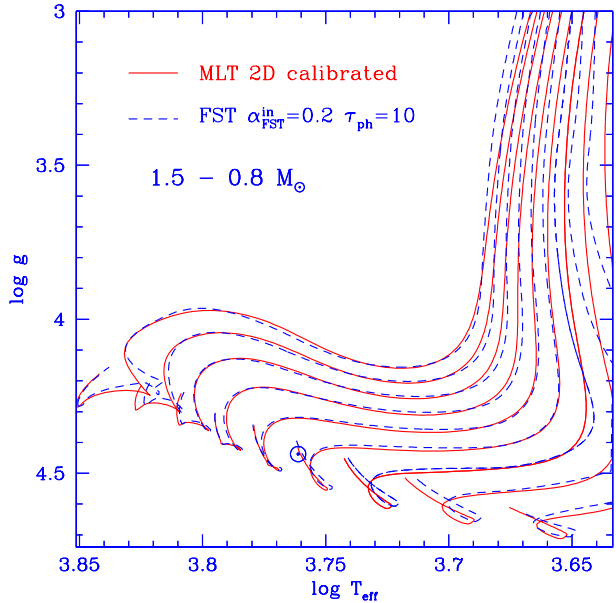}
\end{center}
\vspace{-0.5cm}
\caption{Evolutionary tracks using FST in the $\log T_{\rm eff}$ vs. log~g 
plane (solid line; non--grey models with  $\tau _{\rm ph}=10$ by Montalb\'an et 
al.~(2004a)  and 2D calibrated MLT (dashed line).}
\label{fig1}
\end{figure}

An additional advantage of the procedure of computing \alfa2d models is that 
the entropy in the adiabatic convection region is  independently determined and
is not affected by the ambiguity and the uncertainties deriving from  the use of several
different physics in the over--adiabatic region of the star.
 So, standard non--grey MLT models are not necessarily  equivalent to MLT-2D/3D ones even
 if the numerical 
  value of $\alpha_{\rm MLT}$  parameter used in the stellar 
 structure computations is in both cases the same or of the same order.
 We must indeed 
be careful when looking at the exact meaning of published results. 
 For  instance, the \alfa2d models will {\it not} be similar to the popular set of  
non--grey models having ``$\alpha_{\rm MLT}=1.9$"  by \cite{baraffe98}. In fact, these models 
have indeed $\alpha_{\rm MLT}=\alpha_{\rm MLT}^{\rm in}=1.9$ in the computation of the internal 
structure, but the atmosphere down to $\tau _{\rm ph}=100$, i.e. to the match 
point with with the interior, is computed with $\alpha_{\rm MLT}^{\rm atm}=1$. It has been 
often pointed out that the PMS tracks by \cite{baraffe98}  match better 
observations  since they have lower T$_{\rm eff}$\ than other 
computations (e.g. Hillenbrand \& White 2004) and, sometimes, that has been 
attributed to the improved  opacity in the adopted NextGen atmospheres (Allard 
\& Hauschidt, 1997, hereinafter AH97). On the contrary, for mass values close to the 
solar mass, a large part of this result is simply due  to the very low 
efficiency of convection in the most superadiabatic part of the star, namely the 
atmosphere down to $\tau _{\rm ph}=100$ (Montalb\'an et al.~2004a; D'Antona \& 
Montalb\'an 2003). The calibration of $\alpha_{\rm MLT}$ in \cite{ludwig1999}, whose zero 
point must be obtained by reproducing the solar radius, is valid for grey 
models, so the same $\alpha_{\rm MLT}$ value is used in the whole structure, and the results 
are less ambiguous and easier to be interpreted.

\section{MLT--$\,\alpha^{\rm 2D}$ models}

The MLT--$\,\alpha^{\rm 2D}$ stellar models have been computed with the code ATON2.0 (Ventura 
et al.~1998a), with grey boundary conditions, and the MLT option for convection. 
\cite{ludwig1999}  suggest, in the context of stellar evolutionary models,
to calibrate $\alpha_{\rm MLT}^0$  with the present Sun, and to use its 
ratio to $\alpha_{\rm MLT}^{\rm 2D}(\odot)$ as  scaling factor for 
the function  $\alpha_{\rm MLT}^{\rm 2D}$(\teff,\logg).
So, we  used the value  $\alpha_{\rm MLT}^0=1.6$ (obtained from the grey MLT Sun calibration 
with ATON2.0) to scale the analytical fits of  $\alpha_{\rm MLT}$(\teff,\logg) 
by  \cite{ludwig1999}. 
These  ``solar-calibrated" analytical fits were introduced in ATON2.0 to allow for
a continuous  variation  of the $\alpha_{\rm MLT}$ parameter  as the star evolves
in the HR diagram. 

The adopted helium mass fraction 
is Y=0.28 and the metal mass fraction is Z=0.02. The other physical inputs are 
the same as in  \cite{josi2004}\footnote{The solar mixture used in the
opacity and equation of state tables is that from Grevesse \& Noels 1993}.  In Fig.~\ref{fig1} 
we superimpose the \alfa2d\ tracks (dashed--lines) and the  non--grey FST models 
by \cite{josi2004} in the plane log~g vs. \teff. The latter tracks 
are computed by using the \cite{heiter} boundary conditions, and adopt  FST 
convection, both in the atmosphere and in the interior.  The two 
sets of results, in the range 0.8 -- 1.5\msun\,  are very similar. Since 
stellar radii (and therefore the \teff's) are determined by the efficiency of 
convection in the over-adiabatic region, we can conclude that the global 
efficiency of convection in  \alfa2d\ and FST models is very similar. This 
allows the use of FST models as proxies of 2D RHD models, also for different chemistry.
So, we have computed low metallicity FST models with chemical and
convection  parameters given by the new solar calibration, that is: 
 Y=0.2495, Z=0.01305 and  the FST $\alpha_{\rm FST}$ parameter equal to 0.117.

\begin{figure}
\begin{center}
\includegraphics[width=.48\textwidth]{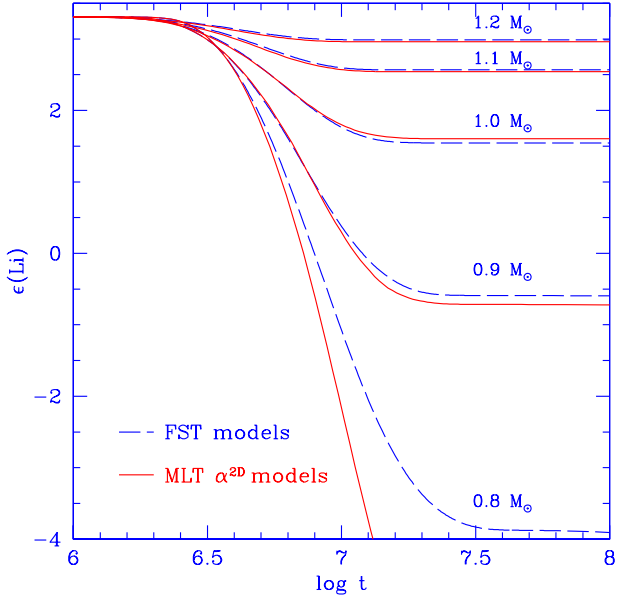}
\end{center}
\vspace{-0.5cm}
\caption{Lithium evolution for models computed with 2D$\alpha$ calibrated MLT 
(solid lines) and complete (non grey) FST models (dashed lines)}
\label{figLimasa}
\end{figure}

\begin{figure}
\begin{center}
\includegraphics[width=.48\textwidth]{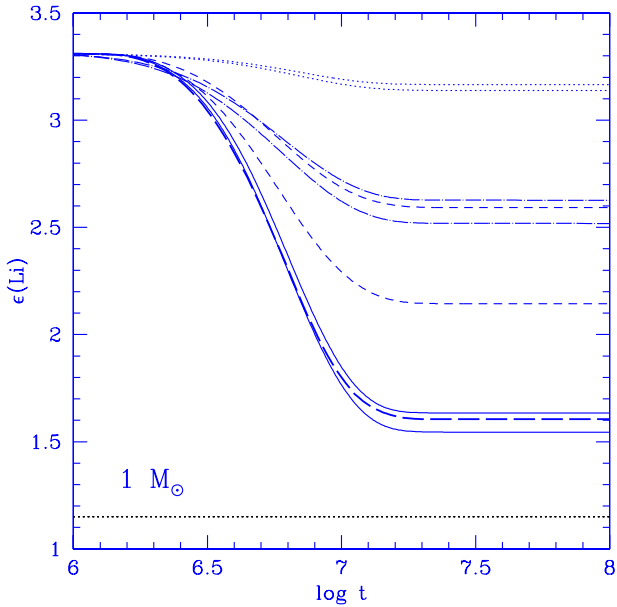}
\end{center}
\vspace{-0.5cm}
\caption{Lithium evolution for the solar mass with different assumptions about 
convection and model atmospheres. The dotted line at bottom represents today's 
solar lithium abundance. MLT models with AH97 model atmospheres down to $\tau 
_{\rm ph}=10$\ and 100 are shown dotted for $\alpha _{\rm in}=1$\ and dash--
dotted for $\alpha _{\rm in}=1.9$. The Montalb\'an et al.~(2004a) MLT models 
with Heiter et al.~(2002) atmospheres down to $\tau _{\rm ph}=10$ (lower) and 
100 (upper) are dashed; The continuous lines show the non--grey  FST models for 
$\tau _{\rm ph}=10$\ and 100, and, in between, the long dashed line shows the 
MLT--$\,\alpha^{\rm 2D}$ model.  }
\label{fig4}
\end{figure}

\subsection{Pre--MS Lithium depletion in MLT--$\,\alpha^{\rm 2D}$  models}

Lithium is a light element that is  burned by the nuclear reaction 
$^7{\rm Li}(p,^4{\rm He})^4{\rm He}$ at a relatively low temperature (T$_{\rm Li}\sim 2.5\,10^6$~K).
 The PMS lithium  depletion rate strongly depends  on temperature and density 
at the bottom of the convective region.
Lithium abundances detected in young open clusters, such as the Pleiades
and $\alpha$Per \citep[e.g.][]{soderblom1993}, 
indicate that solar mass stars do not deplete a significant fraction lithium  during their PMS. 
That implies  structures of these stars with a 
temperature at the base of their convective envelope  smaller than T$_{\rm Li}$. 

It is generally accepted (Gough \& Weiss 1976; Christensen-Darlsgaard 1997) that
the overall late-type stars structure does not depend on the details of the treatment
of over-adiabatic regions. Hence, the radius is determined by the value of the entropy  jump
between the photosphere and the deep adiabatic regions.
Christensen-Dalsgaard (1997) pointed out that, for the Sun, the depth
of the convective zone is almost insensitive to  changes in the adiabat of the
convective zone, and that the changes in the surface radius are due only to 
the changes of the radius of the bottom of the convective zone.
He also pointed out that the same should probably occur for other stars, if the 
derivatives of opacity with respect to temperature and density are of the same order than
in the Sun.
The similarity between  FST and MLT--$\,\alpha^{\rm 2D}$--track locations suggests that the 
``average efficiency''
of convection is quite close in both series of models, and hence 
we expect that also the lithium depletion  in the MLT--$\,\alpha^{\rm 2D}$ models should be similar to 
that of  the FST ones. In Fig.~\ref{figLimasa}   we plot the lithium depletion as a function of time for
masses between 0.8 and 1.2~\msun. In these computations the initial lithium
abundance was taken as log N(Li)=3.31, i.e. the solar system abundance given by \cite{ag1989}\footnote{The current adopted value for the meteoritic Li abundance is $3.25\pm0.06$ \citep{asplund05b}, but the small difference with respect to 
Anders \& Grevesse (1990) does not  significantly affect our results}.
Both the  \alfa2d models and the  non--grey FST models show a depletion of the order of 
$\sim 1.7$~dex for the 1~\msun\ evolutionary track. 
Thus the non--grey FST models provide a description of the stellar structure 
very similar to the RHD 2D models, also with respect to  lithium depletion.

Generally, local convection models are calibrated by requiring that the free 
parameter(s)  reproduces the solar radius at the solar age. The MLT models 
satisfying this constraint predict, however, a pre--MS lithium depletion which 
is not compatible with that observed in young open clusters 
\citep{dm2003}. In Fig.~\ref{fig4} we plot the lithium vs. time evolution in some 
1\msun\ models by \cite{josi2004} and in the \alfa2d models. As \cite{dm2003} 
remarked, the models adopting AH97 atmospheres and $\alpha_{\rm MLT}^{\rm in}$=1 deplete lithium in 
pre--MS by only 0.15~dex. The same models with $\alpha_{\rm MLT}^{\rm in}$=1.9 deplete 0.7~dex if the 
matching point is \taufot=100, but more than 1~dex for \taufot=10. 
We stress again that this result is only due  to the use of a {\bf less efficient
 convection} when the match with the atmosphere ($\alpha_{\rm MLT}^{\rm atm}$=1) 
 is done at  \taufot=100, 
even if the $\alpha_{\rm MLT}$ parameter chosen for the interior, in order to fit the solar 
radius, is much larger (for instance 1.9). As it was shown in \cite{josi2004}
and recalled in Sect.~1, the choice of the
optical depth  at which we match the two parts of the stellar structure, is not without consequences.
It implies an uncertainty of almost 200~K for the \teff\ of 1~\msun\ PMS.
{\it The optical depth at which the external layers change from adiabatic to over-adiabatic 
stratification depends on \teff, \logg, and chemical composition as well. 
To match an $\alpha_{\rm MLT}=1$ structure with  an $\alpha_{\rm MLT}$=1.9 one at  fixed
\taufot\  along the whole evolution is equivalent to decrease the average 
efficiency of convection in a complicate, hidden and unjustified way}.

Li abundance in young clusters suggests that, if no other physical process is affecting the
location of the base of the convective zone, the PMS-tracks should be cooler than solar 
calibrated convection predicts. This evidence should be used to look for the shortcomings 
of our stellar models.

\subsection{Pre--MS binaries}
\begin{figure}
\begin{center}
\vspace*{-0.9cm}
\includegraphics[width=.48\textwidth]{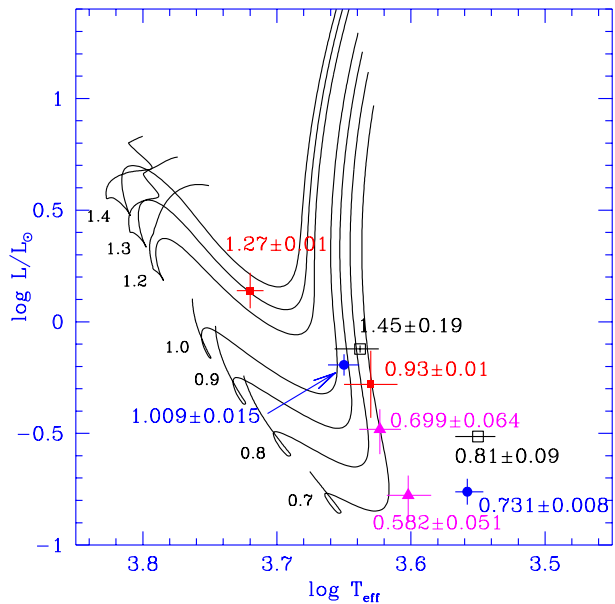}
\end{center}
\vspace*{-2.cm}
\caption{The location in the HR diagram of the MLT--$\,\alpha^{\rm 2D}$ tracks is shown 
together with four binaries with well determined masses, labeled in the 
figure. Full squares (red): RXJ 0529.4+0041 \citep{covino2004}; 
full circles (blue): V1174 Ori \citep{stassun}; open squares (black): NTT 
045251+3016 \citep{steffen}; full triangles (magenta): HD~98800~B \citep{boden} }
\label{fig3}
\end{figure}

The location of pre--MS binaries in the HR diagram is a powerful way of 
constraining  stellar models, but unfortunately its indications are, by now, 
still ambiguous, because of the dearth of systems with well known  masses and 
and good atmospheric parameters. For some of these binaries, the location is 
consistent with models having high convection efficiency, but for others this 
is not the case. 
In Figure~\ref{fig3} we compare the \alfa2d tracks with the parameters derived from
the observations of  four binary systems: 
 RXJ 0529.4+0041  (according to the most recent determination by 
\cite{covino2004}), V1174 Ori (Stassun et al. 2004), NTT 045251+3016 (Steffen 
et al. 2001) and HD~98800~B \citep{boden}. 
A detailed analysis is out of the scope of this paper, but it is 
evident that the agreement is reasonably good for most of the primaries but much
less satisfactory for the secondary,  i.e. less massive, components. 
The V1174~Ori 
primary mass (1\msun) is consistent with the corresponding \alfa2d\ track, as well as the
primary masses of RXJ 0529.4+0041 (1.3\msun) and HD~98800~B (0.699\msun), but 
that of NTT 045251+3016 is rather different from the expected value
(1.45$\pm$0.19\msun) is 
\footnote{the same \cite{steffen} remark that the 
location of the primary in this system is 3$\sigma$\ away from the 
\cite{dm1994} grey--FST tracks, which are hotter than the \alfa2d\ tracks by 
$\sim$270~K.}. 
Besides, at least two secondaries seem to be too cool for the \alfa2d\ tracks,  
and one could perhaps guess  a systematic behavior, in the sense that the low 
mass PMS stars seem to be cooler and to have a larger than expected radius. 

This recalls what happens  for the few known late--type low--mass MS binary components.  
High quality observations (eclipsing binaries and/or interferometric measurements)
revealed a meaningful discrepancy between  stellar parameters derived from observations and those
based on theoretical models: the theoretical stellar radii seem to be  be underestimated by 10-20\%, 
and the effective temperatures overestimated by 5\% (200~K) (e.g. Ribas 2006; Torres et al. 2005;
and references therein).
An accepted, but not definitely proved, explanation is  that these larger than expected 
stellar radii are related to  stellar activity. In fact, the magnetic fields associated 
with stellar activity can decrease the efficiency of convection (Gough \& Tayler 1966; 
Stein et al. 1992 and references therein) and as a consequence
the stellar radius must grow to transport the same quantity of energy.
Besides, the decrease in effective temperature could be due to spots on the
stellar surface. 
In a very recent paper, Torres et al.~(2005) have shown that the discrepancy between theory
and observation can also appear for stars quite similar to  the Sun.
In their study of the eclipsing binary V1061 Cygni with 
$M_{\rm Aa}=1.282$\msun \ and $M_{\rm Ab}=0.9315$\msun, 
they conclude: {\it ``Current stellar evolution models that use a mixing 
length parameter $\alpha_{ML}$ appropriate for the Sun agree well with the 
properties of the primary, but show a very large divergence in the radius of 
the secondary, in the sense that the predicted values are $\sim 10$\% smaller 
than observed. In addition, the temperature is cooler than predicted by some 200~K ... 
for a star only 7\% less massive than the Sun''}.
They suggest
 that there must be  a relation among  activity level,  decrease of
\teff\ and  
underestimation of the stellar radii  by the standard stellar models 
(calibrated by matching the Sun).

\section{FST models with revised solar metallicity}
We have  considered, until now,
 \alfa2d\ models computed for the ``standard" solar 
metallicity Z=0.02, but what will happen if we decrease Z to $\sim$0.013, 
as suggested by the re--analysis of the solar spectrum by \cite{asplund2004}? 
The lithium depletion would probably be reduced, without affecting too much the 
pre--MS tracks location, as we still must calibrate convection in order to 
reproduce the solar location. 
RHD 2D models are not 
available for this revised metallicity, nevertheless, we can compute models 
with FST convection, and make the hypothesis that, for the reduced solar 
metallicity, they provide similar results to \alfa2d\ models, as they do for 
the standard solar metallicity.
In fact, the effective temperature in the Hayashi track corresponding to the
solar model, calibrated with the new solar metallicity, is only 0.3\% lower than that 
for the standard solar metallicity. However, the temperature and density at the bottom of
the convective zone of these PMS  models, have changed by 6 and 12~\% respectively.
As a consequence,  lithium has been depleted only by 0.5~dex.
We recall, however, that the solar model with new solar metallicity is not able to 
fit either the depth of the convective zone, or the sound speed profile inside the Sun.

Since the new solar abundances were published, several teams  tried to recover the
good match between the helioseismic Sun and the standard solar model. They  proposed to
increase the 
opacity, the microscopic diffusion coefficients, or a combination of
both (e.g. Basu \& Antia 2004, Montalb\'an et al. 2004b, Guzik et al. 2005).
We recall, that none of the proposed solutions is fully satisfactory. Nevertheless, 
for what concerns the lithium problem, whatever solution be adopted, it should be able to 
change the bottom of the convective zone during MS without changing the PMS models.
If the discrepancy between the standard solar model and helioseismology could be solved 
only changing the microscopic diffusion (not reliable
 at the moment, see for instance Guzik et al. 2005),
since this process is very slow, perhaps we could keep a low lithium depletion. 
For other processes, we should justify why  they  work during MS and not during solar PMS.
Basu \& Antia (2005) and Bahcall et al. (2005)  proposed the solution of increasing the Ne abundance
(not directly observed in the solar spectrum) by a factor $\sim 3.5$. This suggestion has been
supported by Drake \& Testa (2005) which, on the basis of Ne abundance determination in active stars, 
argue that the solar Ne could be 2.5 times larger than the value adopted for the solar mixtures. 
Without entering in  the controversy  about these new Ne abundance determination \citep[see e.g. ][]{sch05,young2005,asplund05c}, 
we would like to recall that Ne is, with O,  one of the main contributors to opacity at the
bottom of the solar convective zone (close to $T_{\rm Li}$). The solution proposed by Basu \& Antia (2005)
implies to increase the opacity by replacing the decrease of O abundance in the new solar mixture (Asplund et al.
2004) by the increase of Ne one. A direct comparison between  OPAL opacity tables, with GN93 mixture, and
a solar mixture with Ne increased by a factor 3.5 yields that the opacity from this latter table, for
the temperature and density typical of PMS stars, is even larger than for GN93 mixture. 
As a consequence, one can predict a deeper convective region in the PMS and, therefore, a lithium depletion
 even  larger  than in the "old" standard solar model.

\section{Conclusions}

The computation of MLT pre--MS tracks in which the $\alpha_{\rm MLT}$ parameter is 
calibrated on 2D RHD models (\alfa2d\ tracks) shows that these tracks deplete 
too much lithium, at variance with the observations of stars in young open 
clusters. The \alfa2d\ tracks are similar, in HR diagram location and lithium 
depletion, to the FST non--grey tracks by \cite{josi2004}. Thus we compute FST 
non--grey models, correponding to the new solar metallicity determined by 
\cite{asplund2004}, as proxies of possible 2D RHD models for the reduced 
solar metallicity. The lithium depletion in 1~\msun\ model  has been reduced from 1.7~dex 
(with the ``old" solar metallicity) to $\sim$~0.5~dex.  We know, however, that such a low metalllicity 
is not able to reproduce other, well established,
properties of the Sun. Hence, much more work shall be done, analyzing all the  consequences of the
new solar metallicity, before adopting the new abundances  as solution of the Li problem.

The problem of lithium is therefore not solved in the framework of 
the numerical simulations of convection. On the contrary, 2D and 3D 
numerical simulations imply even more efficient convection in the 
PMS, while  lithium depletion in young clusters  and PMS binaries studied here suggest a
lower efficiency. 
 We conclude that convection in pre--MS must be less efficient than 
what is suggested by 2D RHD models.
It is still possible that the description 
of pre--MS convection requires the introduction of a second parameter --linked 
to the stellar rotation and magnetic field, as we have suggested in the past 
(Ventura et al.~1998b; D'Antona et al.~2000).

The data on masses and radii coming from 
high quality observations of eclipsing binaries and/or interferometry
of late type stars   agree only in part with the \alfa2d models.
So, while some observations (stellar parameters of binaries 
and lithium abundances in young clusters) require  a
low efficiency of convection in the late-type
domain of the HR diagram, numerical simulations of convection (Ludwig et al. 1999; 
Trampedach et al. 1999; Ludwig et al. 2000)  show the opposite: 
an equivalent value of the $\alpha_{\rm MLT}$ parameter that increases as
effective temperature decreases. 

The abovementioned observational results, together with the high activity level that PMS-stars
can show (see e.g. Tayler 1987), seems to support  the suggestion 
that 
that other physical processes (rotation, magnetic fields) affect the efficiency of 
convection in  late type stars. 
\section*{Acknowledgments}
The authors thank the referee for his constructive and useful comments.
JM acknowledges financial support from the Prodex-ESA Contract 15448/01/NL/Sfe(IC)

\end{document}